\documentclass[aps,prl,twocolumn,showpacs,showkeys,preprintnumbers,floatfix,nofootinbib,superscriptaddress,longbibliography]{revtex4}
\usepackage{amsfonts} % AMS
\usepackage{amssymb} % AMS
\usepackage{amsmath} % AMS
\usepackage{graphicx} % Include figure files
\usepackage{subfigure} % Include figure files
\usepackage{array} % array
\usepackage{dcolumn} % Align table columns on decimal point
\usepackage{latexsym} % latex symbols
\usepackage{longtable} % long tables
\usepackage{bbold}
\usepackage{xcolor}
\usepackage{multirow}
\usepackage{rotating}
\usepackage{comment}
\usepackage{zref-xr}
\usepackage{zref-user}
\usepackage{slashed}
\usepackage{ulem}

\newcommand{\bea}{\begin{eqnarray}}
\newcommand{\eea}{\end{eqnarray}}
\newcommand{\be}{\begin{equation}}
\newcommand{\ee}{\end{equation}}

\newcommand{\slashpi}{\protect{\slash\hspace{-0.5em}\pi}}

\usepackage{textgreek}
\usepackage[unicode]{hyperref} % hypertext links
\hypersetup{
    colorlinks=true,       % false: boxed links; true: colored links
    linkcolor=blue,          % color of internal links
    citecolor=blue,        % color of links to bibliography
    filecolor=blue,      % color of file links
    urlcolor=blue           % color of external links
}

\definecolor{green}{rgb}{0.1, 0.8, 0.1}

\usepackage{textgreek}
\usepackage[unicode]{hyperref} % hypertext links

\begin{document}

%%%

%\title{Percent level hadronic isospin breaking corrections to neutron $\beta$-decay}
%\title{Radiative correction to neutron decay in Heavy Baryon Chiral Perturbation Theory}
%\title{Radiative correction to neutron decay in hadronic effective field theory}
\title{Pion-induced radiative corrections to neutron beta-decay}

\author{Vincenzo~Cirigliano}
\email{cirigliano@lanl.gov}
\affiliation{Los Alamos National Laboratory, Theoretical Division T-2, Los Alamos, NM 87545, USA}
\affiliation{Institute for Nuclear Theory, University of Washington, Seattle WA 98195-1550}

\author{Jordy~de~Vries}
\email{j.devries4@uva.nl}
\affiliation{Institute for Theoretical Physics Amsterdam and Delta Institute for Theoretical Physics,
University of Amsterdam, Science Park 904, 1098 XH Amsterdam, The Netherlands}
\affiliation{Nikhef, Theory Group, Science Park 105, 1098 XG, Amsterdam, The Netherlands}

\author{Leendert~Hayen}
\email{lmhayen@ncsu.edu}
\affiliation{Department of Physics, North Carolina State University, Raleigh, North Carolina 27695, USA}
\affiliation{Triangle Universities Nuclear Laboratory, Durham, North Carolina 27708, USA}

\author{Emanuele~Mereghetti}
\email{emereghetti@lanl.gov}
\affiliation{Los Alamos National Laboratory, Theoretical Division T-2, Los Alamos, NM 87545, USA}

\author{Andr\'e~Walker-Loud}
\email{walkloud@lbl.gov}
\affiliation{Nuclear Science Division, Lawrence Berkeley National Laboratory, Berkeley, CA 94720, USA}

\preprint{LA-UR-21-31960}
\preprint{INT-PUB-22-005}

\begin{abstract}
We compute the electromagnetic
%radiative QED
 corrections to neutron beta decay using a low-energy hadronic effective field theory.
%The relevant scales of the problem, i.e. $q_{\rm ext}  \sim  O(m_e, m_n - m_p)$, $m_\pi$ and $\Lambda_\chi  \sim O(m_n) \sim  O(4 \pi F_\pi)$ are used to form small parameters
%$\alpha = e^2/(4 \pi)$,
%$\epsilon_{\rm recoil} \sim q_{\rm ext} /m_n$, $\epsilon_{\slashpi} \sim q_{\rm ext}/m_\pi$, and $\epsilon_\chi \sim m_\pi/\Lambda_\chi$, that are used to systematically determine the amplitude and its theoretical uncertainty.
%We work to next-to-leading order in the pion-full theory, which is $\mathrm{O}(G_F  \alpha \epsilon_{\chi})$ and match the result to the lower-energy pion-less theory at $\mathrm{O}(G_F  \alpha \epsilon_{\slashpi})$.
%In the process we identify previously overlooked electromagnetic corrections to $g_A/g_V$ of percent-level size that are proportional to the pion electromagnetic mass splitting --
%the result depends on calculable chiral loop contributions as well as
%a linear combination of currently unknown low-energy constants.
%We comment on the implications of our results for
%the comparison of the experimentally measured $g_A/g_V$ with the corresponding lattice QCD calculations
%and for   the extraction of bounds on right-handed currents from beta  decays.
We identify and compute new radiative corrections arising from virtual pions that were missed in previous studies. The largest correction is a percent-level shift in the axial charge of the nucleon proportional to the electromagnetic part of the pion-mass splitting. Smaller corrections, comparable to anticipated experimental precision, impact the $\beta$-$\nu$ angular correlations and the $\beta$-asymmetry.  We comment on implications of our results for the comparison of the experimentally measured  axial charge with first-principle computations using lattice QCD  and  on the potential of $\beta$-decay experiments to constrain beyond-the-Standard-Model interactions.

%-------------------------------------------------------------------------------
% OLD ABSTRACT BELOW
\iffalse
We perform the calculation of  radiative corrections to neutron decay in
a low-energy hadronic effective field theory, keeping track of all the relevant scales
in the problem, i.e. $\Lambda_\chi  \sim O(m_n) \sim  O(4 \pi F_\pi)$,
$m_\pi$, and $q_{\rm ext}  \sim  O(m_e, m_n - m_p)$.
We  expand  the neutron decay amplitude in terms of the small parameters
$\alpha = e^2/(4 \pi)$,  $\epsilon_{\rm recoil} \sim q_{\rm ext} /m_n$,
$\epsilon_{\slashpi} \sim q_{\rm ext}/m_\pi$, and $\epsilon_\chi \sim m_\pi/\Lambda_\chi$.
We set up the analysis within Heavy Baryon Chiral Perturbation Theory extended to include
virtual photons and leptons and match it to a  pion-less theory to order  $G_F \epsilon_{\rm recoil}$,
$G_F \alpha$,  $G_F  \alpha \epsilon_{\chi}$,  and  $G_F  \alpha \epsilon_{\slashpi}$.
In the process we identify previously overlooked electromagnetic corrections to $g_A/g_V$ of percent-level size
at $O(G_F \alpha)$ and $O(G_F \alpha \epsilon_\chi)$,
proportional to the pion electromagnetic mass splitting.
These  depend on calculable chiral loop contributions as well as
a linear combination of currently unknown  low-energy constants (LECs).
We outline strategies to determine these LECs  from lattice QCD calculations.
We briefly discuss the implications of our results for
the comparison of the experimentally measured $g_A/g_V$ with the corresponding lattice QCD calculations
and for   the extraction of bounds on right-handed currents from beta  decays.
\fi
\end{abstract}
\maketitle
%

%\section{Introduction}

{\it Introduction} ---
High-precision measurements of low-energy processes, such as $\beta$ decays of mesons, neutron, and nuclei, probe the existence of new physics at very high energy scales through quantum fluctuations. Recent developments in the study of $\beta$ decay rates at the sub-\% level \cite{Seng:2018yzq,Seng:2018qru,Czarnecki:2019mwq,Shiells:2020fqp,Hardy:2020qwl} have led to a 3-5$\sigma$ tension with the Standard Model (SM) interpretation in terms of the unitary Cabibbo-Kobayashi-Maskawa (CKM) quark mixing matrix \cite{Hardy:2020qwl,ParticleDataGroup:2020ssz}. Further, global analyses of $\beta$ decay observables \cite{Falkowski:2020pma,Gonzalez-Alonso:2018omy} have highlighted additional avenues for $\beta$ decays to probe physics beyond the Standard Model (BSM) at the multi-TeV scale, such as the comparison of the experimentally extracted
%strength of the Gamow-Teller coupling ($g_A$ for neutron decay)
weak axial charge, $g_A$,
with precise lattice Quantum ChromoDynamics (QCD) calculations \cite{Bhattacharya:2011qm,Alioli:2017ces,Chang:2018uxx}. This test is a unique and sensitive probe of BSM right-handed charged currents.%.quark to $W$ boson right-handed couplings.

Given the expected improvements in experimental precision in the next few years \cite{Cirgiliano:2019nyn, Pocanic2009, Dubbers2008}, a necessary condition to use neutron decay as probe of BSM physics is to have high-precision calculations {\it within the SM}, including sub-\% level recoil and radiative corrections with controlled uncertainties. These prospects have spurred new theoretical activity, which has focused first on radiative corrections to the strength of the Fermi transition (vector coupling) \cite{Seng:2018yzq,Seng:2018qru,Czarnecki:2019mwq,Shiells:2020fqp}, and more recently on the corrections to the Gamow-Teller (axial) coupling~\cite{Hayen:2020cxh,Gorchtein:2021fce}. These recent studies are all rooted in the current algebra approach developed in the sixties and seventies \cite{Sirlin:1967zza, Sirlin:1977sv}, combined with the novel use of dispersive techniques.

In principle, lattice QCD can be used to compute the full Standard Model $n\rightarrow pe\bar{\nu}$ decay amplitude including radiative QED corrections, similar to the determination of the leptonic pion decay rate~\cite{Carrasco:2015xwa,Giusti:2017dwk}.
However, it will be some years before these calculations reach sufficient precision.
% exist, and more still before they are achieve sufficient precision for these Standard Model tests:
Currently, lattice QCD calculations are carried out in the isospin limit.
The global average determination of $g_A$ carries a 2.2\% uncertainty~\cite{Aoki:2021kgd} with one result achieving a 0.74\% uncertainty~\cite{Chang:2018uxx,Walker-Loud:2019cif}. The PDG average value, on the other hand, has an 0.1\% uncertainty~\cite{ParticleDataGroup:2020ssz} with the most precise experiment having an 0.035\% uncertainty~\cite{Markisch:2018ndu}.

In this work, we  perform a systematic study of neutron decay using effective field theory (EFT).
 We compute new  structure-dependent electromagnetic corrections originating at the pion mass scale,
including effects of $\mathcal{O}(\alpha)$ and $\mathcal O(\alpha m_\pi/m_N)$, with $\alpha = e^2/4\pi$ the fine-structure constant and $m_\pi (m_N)$ the pion (nucleon) mass.  By doing so we uncover new {\it percent-level} electromagnetic corrections to the axial coupling $g_A$, which were missed both in the only other neutron $\beta$ decay EFT analysis ~\cite{Ando:2004rk} and recent dispersive treatments~\cite{Hayen:2020cxh,Gorchtein:2021fce}.

{\it Neutron decay from the Standard Model ---}
The energy release in neutron decay is roughly the mass splitting of the neutron and proton, i.e. $q_\mathrm{ext} \sim m_n-m_p \sim 1$ MeV, which is significantly smaller than the nucleon mass. The energy scale of nucleon structure corrections, on the other hand, is related to the pion mass, so that $m_N \gg m_\pi \gg m_n - m_p$. Large scale separations, such as these, make for ideal systems for an EFT description.

As a consequence, corrections to neutron $\beta$ decay can be parametrized in terms of two small parameters: ($i$) $\epsilon_{\rm recoil} = q_{\rm ext}/m_N \sim 0.1\%$ which characterizes small kinetic corrections; ($ii$) $\epsilon_{\slashpi}= q_{\rm ext}/m_\pi \sim 1\%$, which characterizes nucleon structure corrections dominated by radiative pion contributions. At these relatively low energies, the decay amplitude can be  described by a non-relativistic Lagrangian density (see also Refs.~\cite{Ando:2004rk,Falkowski:2021vdg})
\begin{align}\label{eq:Lpiless}
 \mathcal L_{\slashpi} &=
 - \sqrt{2} G_F V_{ud} \, \bigg[
 \bar e \gamma_\mu P_L \nu_e
  \bigg(
 \bar N \left(g_V v_\mu - 2 g_A S_\mu \right) \tau^+ N
 \nonumber \\&
  +\frac{i}{2 m_N} \bar N
 (v^\mu v^\nu - g^{\mu \nu} - 2 g_A v^\mu S^\nu) ( \overleftarrow  \partial - \overrightarrow \partial)_\nu \tau^+ N
 \bigg)
 %\ \bar e \gamma_\mu P_L \nu_e
 \nonumber \\&
+  \frac{i c_T m_e}{m_N} \bar N \left( S^\mu v^\nu - S^\nu v^\mu\right) \tau^+ N\,  \left( \bar e \sigma_{\mu \nu} P_L \nu \right)\,
 \nonumber \\&
 %- \sqrt{2} G_F V_{ud}
+\frac{i \mu_{\rm weak}}{m_N}  \bar N [S^\mu, S^\nu] \tau^+ N\,  \partial_\nu \left( \bar e \gamma_\mu P_L \nu \right)  \bigg] + \dots
% &+ & \frac{i}{2 m_N} \bar N
% (v^\mu v^\nu - g^{\mu \nu} - 2 g_A v^\mu S^\nu) ( \overleftarrow  \partial - \overrightarrow \partial)_\nu \tau^+ N \ \bar e \gamma_\mu P_L \nu_e
% \nonumber \\
  %- \sqrt{2} G_F V_{ud}
%&+ &
 % & & + \frac{i g_T m_e}{4 m_N} \bar N \left( S^\mu v^\nu - S^\nu v^\mu\right) \tau^+ N\,  \left( \bar e \sigma_{\mu \nu} P_L \nu \right)
%\nonumber \\ & +& \dots
\end{align}
where pions have been integrated out (hence subscript $\slashpi$), and the ellipsis denote terms not affected by our analysis.
%\emanuele{Do we want to use the nomenclature $\slashpi$EFT? It might carry some baggage. I added $\ldots$ cause it is not clear the above Lagrangian is complete. E.g. the induced scalar and induced axial tensor could in principle be there. Is there an obvious way to count isospin breaking?}
In this expression, $N^T=(p,n)$ is an isodoublet of nucleons, while $v_\mu$ and $S_\mu$ represent the velocity and spin of the nucleon, respectively. The effective vector and axial-vector couplings $g_{V,A}$ are related, as discussed below, to the isovector nucleon vector and axial charges,
%$g_V$ is the isovector nucleon charge and $g_A$ is the isovector axial coupling.
while $\mu_{\rm weak}$ and $c_T$ are the weak magnetic moment and an effective tensor coupling, respectively.
%In principle, these LECs can be measured experimentally, and then used with this Lagrangian, to predict the neutron decay amplitude.
 The Lagrangian \eqref{eq:Lpiless} can be used to compute the differential neutron decay rate and the parameters can then be fitted to data.
%\emanuele{With the Lagrangian \eqref{eq:Lpiless} one can calculate the neutron decay rate and correlation coefficients, and adjust the parameters to reproduce experimental data. }

There are a number of short-comings to this approach. First, by utilizing measured values of $V_{ud} \,g_V$, $g_A/g_V$, $\mu_{\rm weak}$, and $c_T$, we cannot extract fundamental SM parameters nor distinguish SM from BSM contributions to these low-energy constants (LECs). Second, it is not possible to disentangle, for example, how much of $g_A$ arises from isospin symmetric QCD versus
%radiative structure corrections induced by
electromagnetic contributions. Therefore, it is desirable to utilize an EFT Lagrangian which encodes the corrections as functions of the SM parameters, such as the quark masses and the electromagnetic couplings. This is known as chiral perturbation theory ($\chi$PT)~\cite{Gasser:1983yg,Gasser:1984gg}, or specifically for baryons, heavy baryon $\chi$PT (HB$\chi$PT)~\cite{Jenkins:1990jv}. The cost of such a description is the introduction of new scales, $m_\pi$ and $\Lambda_\chi = 4 \pi F_\pi \sim 1$ GeV with $F_\pi \simeq 92.4$ MeV, which form another expansion parameter, $\epsilon_\chi = m_\pi/\Lambda_\chi$, and new operators with potentially undetermined LECs.

Radiative corrections to neutron decay can be organized in a double expansion in $\alpha \epsilon^n_\chi \epsilon^m_\slashpi$.
First, we integrate out the pions and match the $\chi$PT amplitude to the $\slashpi$EFT amplitude, thus determining the quark mass and electromagnetic corrections to effective couplings such as $g_A$. Then, the neutron decay amplitude can be computed with $\slashpi$EFT (with dynamical photons and leptons) while retaining explicit sensitivity to the parameters of the Standard Model.
%by first integrating out pions and matching $\chi$PT onto $\slashpi$EFT,
%and then by computing neutron decay in $\slashpi$EFT.
In our analysis of the decay amplitude we retain terms of $\mathcal O(G_F \epsilon_{\rm recoil})$, known in the literature, $\mathcal O(G_F \alpha)$, where we uncover previously overlooked effects, and terms of $\mathcal O(G_F \alpha \epsilon_{\chi})$ and $\mathcal O(G_F \alpha \epsilon_{\slashpi})$, never before considered in the literature.

\bigskip

{\it $\chi$PT setup for neutron decay --- }
To study radiative corrections to  weak semi-leptonic transitions, we adopt the  HB$\chi$PT framework \cite{Jenkins:1990jv} with dynamical photons \cite{Meissner:1997ii,Muller:1999ww,Gasser:2002am} and leptons,
%~\cite{Ando:2004rk}),,
in analogy with the meson sector~\cite{Knecht:1999ag}. This EFT provides a necessary intermediate step in the analysis of neutron decay, before integrating out pions, and is the starting point for the study of related processes such as muon capture, low-energy neutrino-nucleus scattering,  and nuclear $\beta$ decays, which of course require a non-trivial generalization to multi-nucleon effects.

In $\chi$PT with dynamical photons and leptons,  semileptonic  amplitudes are expanded  in  the Fermi constant $G_F$ (to first order), the electromagnetic fine structure constant $\alpha$, and $\epsilon_\chi$, while keeping all orders in $q_{\rm ext}/m_\pi$, according to Weinberg's power counting~\cite{Weinberg:1978kz,Weinberg:1990rz,Weinberg:1991um}.
%To a given order in the expansion, tree-level and loop amplitudes  contribute, with insertions of vertices from
%effective Lagrangians ordered according to their chiral dimension.
Following standard practice, derivatives ($\partial \sim p$)  and  the  electroweak couplings $e$, $G_F$ are assigned chiral dimension one, while the light quark mass is assigned chiral dimension two ($m_\pi^2 \sim p^2$).
%
%($e^2 G_F p$).
The relevant effective Lagrangians, ordered according to their chiral dimension, are
\begin{subequations}
\bea
{\cal L}_\pi &=& {\cal L}_\pi^{(2)} + ...
\\
{\cal L}_{\pi N} &=& {\cal L}_{\pi N}^{(1)} +  {\cal L}_{\pi N}^{(2)} +   {\cal L}_{\pi N}^{(3)} + ...
\\
{\cal L}_{\rm lept} &\equiv&  {\cal L}_{\rm lept}^{(1)} =
\bar e \left( i  \slashed{\partial} + e \slashed{A} - m_e \right) e  + \bar \nu  i \slashed{\partial} \nu~.
\eea
\end{subequations}
At a given chiral dimension, one can further separate the strong and electromagnetic Lagrangians
\begin{subequations}
\bea
{\cal L}_\pi^{(2)} &=&
{\cal L}_\pi^{p^2}  +
{\cal L}_\pi^{e^2 p^0}
%\\
%{\cal L}_\pi^{(4)} &=&
%{\cal L}_\pi^{p^4}  +
%{\cal L}_\pi^{e^2 p^2}
%+ {\cal L}_{\pi \ell} ^{e^2 p^2}
\\
{\cal L}_{\pi N}^{(1)} &=&
{\cal L}_{\pi N}^{p}
% =  \bar N_v i v \cdot \nabla N_v + g_A \bar N_v S \cdot u N_v
\\
{\cal L}_{\pi N}^{(2)} &=&   {\cal L}_{\pi N}^{p^2} + {\cal L}_{\pi N}^{e^2 p^0}
\\
{\cal L}_{\pi N}^{(3)} &=&   {\cal L}_{\pi N}^{p^3} + {\cal L}_{\pi N}^{e^2 p} + {\cal L}_{\pi N \ell}^{e^2 p} ~\,,
\eea
\end{subequations}
whose explicit forms are given in the Appendix,
where for the first time we present
%We present here for the first time
the effective Lagrangian $ {\cal L}_{\pi N \ell}^{e^2 p}$ that reabsorbs
the divergences from one loop diagrams involving nucleons, photons, and charged leptons.

The leading amplitude ${\cal A}^{G_Fp^0}$ arises from one  insertion of the lowest order Lagrangian ${\cal L}_{\pi N}^{p}$ expanded to first order in the external weak currents %(see Appendix~\ref{sect:appendix}
%for details)
\begin{eqnarray}
{\cal L}_{\pi N}^{p} \supset
%\bar N_v i v \cdot \nabla N_v + g_A \bar N_v S \cdot u N_v
- \sqrt{2}  G_F V_{ud}  \, \bar N  \left( v_\mu - 2 g_A^{(0)}  S_\mu \right)  \tau^+ N \ \bar e \gamma_\mu P_L \nu_e~,
\label{eq:L1}
\end{eqnarray}
where $g_A^{(0)}$ denotes the nucleon axial charge in the chiral limit and in absence of electromagnetic effects.

\begin{figure*}[ht]
 \includegraphics[width=0.9\textwidth]{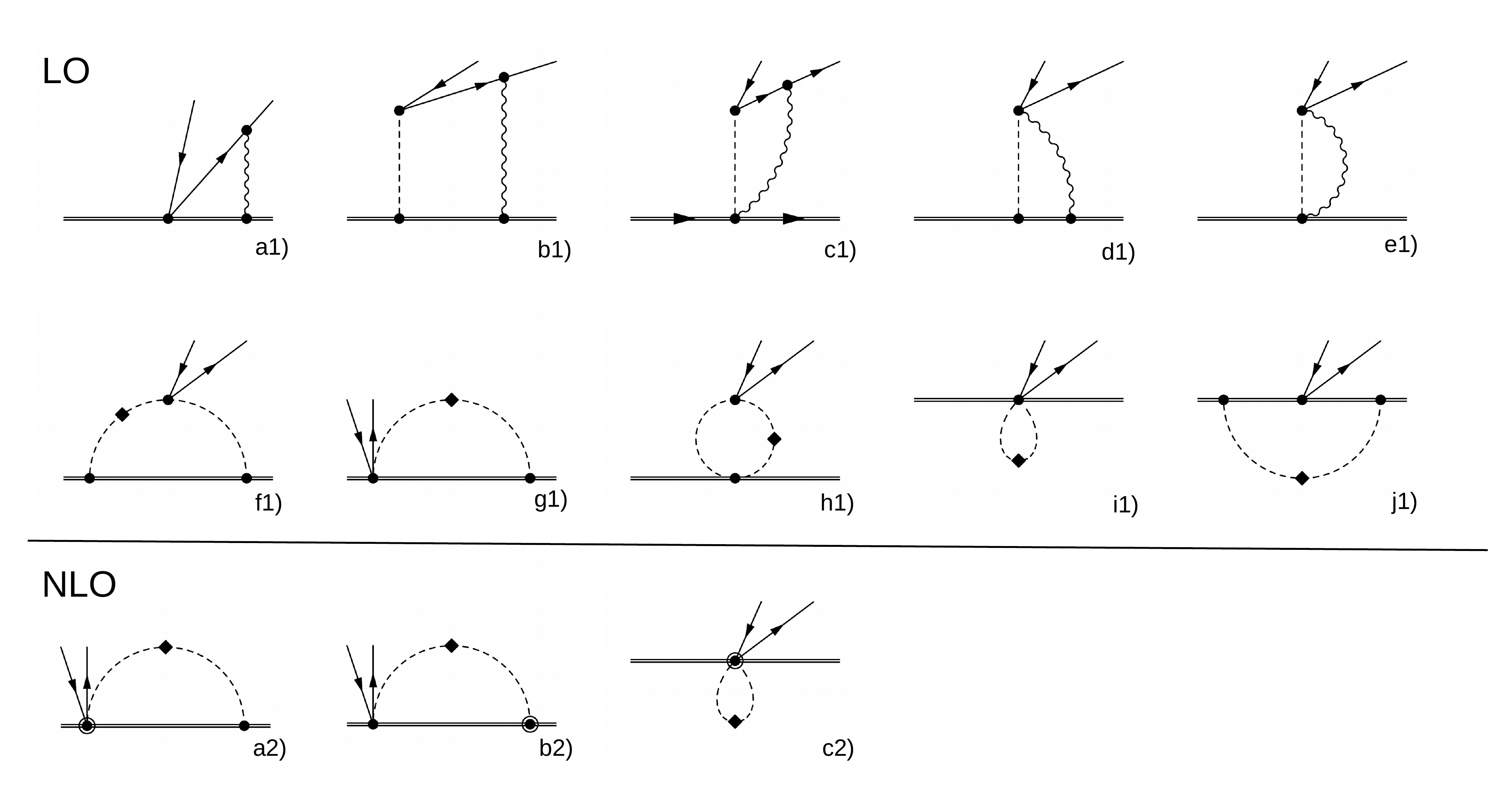}
 \caption{
Diagrams contributing to the matching between $\chi$PT and $\slashpi$EFT at $\mathcal O(\epsilon_\chi^0)$ (upper panel)  and $\mathcal O(\epsilon_\chi)$ (lower panel).
 Single, double, wavy, and dashed lines denote, respectively, leptons, nucleons, photons, and pions. Dots refer to interactions   from the lowest-order chiral Lagrangians ${\cal L}_\pi^{p^2}$ and
 ${\cal L}_{\pi N}^{p}$,  while diamonds represent insertions  of ${\cal L}_\pi^{e^2 p^0}$.
Circled dots denote interactions from the NLO chiral Lagrangian  ${\cal L}_{\pi N}^{p^2}$.
}\label{Fig:DiagLO}
\end{figure*}

To capture electromagnetic corrections to $\mathcal O(G_F \alpha)$,   $\mathcal O(G_F  \alpha \epsilon_{\chi})$, and $\mathcal O(G_F  \alpha \epsilon_{\slashpi})$, we need to compute the neutron decay amplitude to chiral dimension three  (${\cal A}^{e^2 G_Fp^0}$)
%($e^2 G_F$)
and four (${\cal A}^{e^2 G_Fp}$). The former arises from one-loop diagrams  involving virtual nucleons, pions,  photons, and  charged leptons, with vertices from ${\cal L}_{\pi N}^{p}$  and ${\cal L}_{\pi }^{e^2p^0}$ (see Fig. ~\ref{Fig:DiagLO}, upper panel). Here, an important role is played by insertions of
%\jdv{maybe remove the first equality since we have not introduced in the letter the $\mathcal Q_{L,R}$ notation?}
\be
{\cal L}_\pi^{e^2 p^0}
%=  e^2 Z_\pi F^4 \langle {\cal Q}_L^{EM} {\cal Q}_R^{EM} \rangle
=  2 e^2  F_\pi^2 Z_\pi  \pi^+ \pi^-  + O(\pi^4),
\ee
with the LEC $Z_\pi$ fixed by the relation $m_{\pi^\pm}^2 - m_{\pi^0}^2 = 2 e^2  F_\pi^2 Z_\pi$, up to higher-order corrections. Additional contributions arise from tree-level graphs with one insertion of ${\cal L}_{\pi N}^{e^2p}$ or ${\cal L}_{\pi N \ell}^{e^2p}$.
%and any number of insertions from  ${\cal L}_{\pi N}^{p}$  and ${\cal L}_{\pi }^{p^2}$. 
The ${\cal A}^{e^2 G_Fp}$ amplitude, on the other hand,
 % ${\cal A}^{(4)}$
%   is given  by tree-level graphs with one insertion of ${\cal L}_{\pi N}^{(4)}$
%and any number of insertions   from  ${\cal L}_{\pi N}^{(1)}$  and ${\cal L}_{\pi }^{(2)}$;
is a combination of one-loop diagrams %  involving virtual nucleons, pions,  photons, and  charged leptons,
with one vertex from ${\cal L}_{\pi N}^{p^2}$  or ${\cal L}_{\pi N}^{e^2 p^0}$ and any number of  vertices from ${\cal L}_{\pi N}^{(1)}$   and ${\cal L}_{\pi }^{(2)}$ (see Fig.~\ref{Fig:DiagLO}, lower panel).
%Tree-level graphs with insertion of ${\cal L}_{\pi N}^{e^2 p^2}$ do not contribute.

%\bigskip
%\input{pionlessEFT}

\bigskip

%\begin{figure*}
% \includegraphics[width=0.6\textwidth]{MatchingNLO_2.pdf}
% \caption{Diagrams contributing to the matching between $\chi$PT and $\slashpi$EFT at $\mathcal O(\epsilon_\chi)
%$.
% Circled dots denote   interactions from the NLO chiral Lagrangian  ${\cal L}_{\pi N}^{p^2}$.
% %, while diamonds  on a nucleon line represent insertions  of ${\cal L}_{\pi N}^{e^2 p^0}$.
%All other notation is as in Fig.~\ref{Fig:DiagLO}.
%}\label{Fig:DiagNLO}
%\end{figure*}

{\it Matching at $\mathcal O(\alpha)$ and  $O(\alpha \epsilon_\chi)$ -- }
The diagrams contributing to the matching between $\chi$PT and $\slashpi$EFT at $\mathcal O(\epsilon_\chi^0)$
and $\mathcal O(\epsilon_\chi)$ are shown in Fig. \ref{Fig:DiagLO}.
The result of this matching for the leading vector and axial operators is given by
\begin{eqnarray}
 g_{V/A}  &=& g_{V/A}^{(0)} \left[1 + \sum_{n=2}^{\infty} \Delta_{V/A,\chi}^{(n)} +  \frac{\alpha}{2\pi} \sum_{n=0}^{\infty} \Delta^{(n)}_{V/A, \mathrm{em}} \right. \nonumber \\
 & &  \left. +  \left( \frac{m_u - m_d}{\Lambda_\chi}
\right)^{n_{V/A}} \sum_{n=0}^\infty \Delta^{(n)}_{V/A,\delta m}
 \right]\,,
\end{eqnarray}
where $g_V^{(0)}=1$, $\Delta^{(n)}_{\chi, {\rm em}, \delta m} \sim O(\epsilon_\chi^n)$, and $n_A=1$, $n_V=2$~\cite{Behrends:1960nf,Ademollo:1964sr}. Explicit calculation gives $\Delta^{(0),(1)}_{A, \delta m} = 0$ and we do not consider the tiny effect of $\Delta^{(0)}_{V, \delta m} \neq 0$. Concerning the chiral corrections  in the isospin limit,
$\Delta_{V, \chi}^{(n)}$
vanish due to conservation of the vector current, while
$\Delta_{A, \chi}^{(n)}$  have been calculated up to $n=4$ in
Refs.~\cite{Bernard:1992qa,Kambor:1998pi,Bernard:2006te}, and can for our purposes be absorbed into a definition of $g_A$ in the isospin limit, which we denote by $g_A^{\rm QCD}$.

To $O(\alpha \epsilon_\chi^0)$ we consider the diagrams in Fig. \ref{Fig:DiagLO}, upper panel. Diagram $(a1)$ appears in the same form in both EFTs, and thus does not contribute to the matching. An explicit calculation shows that the $\mathcal O(\epsilon_\slashpi^0)$ term of diagrams  $(b1)$ and $(d1)$ and $(c1)$ and $(e1)$ cancels, leaving $\mathcal O(\epsilon_\slashpi)$ corrections discussed below. Diagrams $(g1)$ and $(j1)$ vanish exactly at $\mathcal O(\epsilon_\chi^0)$, while
%, for both axial and vector currents.
$(f1)$, $(h1)$, $(i1)$ contribute to the vector operator only to be cancelled by corrections to the nucleon wavefunction renormalization (WFR) at $q=0$. As a consequence, $g_V$ does not receive loop corrections in the matching between $\chi$PT and $\slashpi$EFT,
instead picking up contributions only from local operators of $\mathcal{O}(e^2p)$ so that $\Delta^{(0)}_{V, \rm em} = \hat{C}_V$.
%so that $g_V$  receives corrections in the matching between $\chi$PT and $\slashpi$EFT only from
%local operators of $O(e^2p)$, not from loops.
By contrast, the axial operator is modified through diagram $(i1)$, the WFR, and local operators of $O(e^2p)$, leading to
%At $q=0$, these corrections lead to the following relation between  $g_A$ and $g_A^{(0)}$
%\begin{eqnarray}
% g_A  &=& g_A^{(0)} \left(1 + \sum_{n=2}^{\infty} \Delta_\chi^{(n)} +  \frac{\alpha}{2\pi} \sum_{n=0}^{\infty} \Delta^{(n)}_{\rm em} \right. \nonumber \\
% & &  \left. +   \frac{(m_u - m_d)}{\Lambda_\chi} \sum_n \Delta^{(n)}_{\delta m}
% \right)\,,
%\end{eqnarray}
%where
%%we denote  by $g_A^{(0)}$  the nucleon axial coupling in the chiral limit  and
%$\Delta^{(n)}_{\chi, {\rm em}, \delta m} \sim O(\epsilon_\chi^n)$.
%
%Finally, the loop diagrams contributing to $\Delta^{(0)}_{\rm em}$ are ultraviolet divergent, and the divergence is
%absorbed by a combination of counterterms denoted by
%$\hat{C}_A(\mu)$
\begin{equation}
 \Delta^{(0)}_{A, \rm em}
 = Z_\pi \left[\frac{1 + 3 g^{(0) 2}_A}{2}
 \left(  \log \frac{\mu^2}{m_\pi^2} -1 \right) - g_A^{(0)2} \right]+ \hat{C}_{A}(\mu)~.
 \label{eq:dAem0}
\end{equation}
%%%%%%%%%%
%\begin{eqnarray}
% \Delta^{(0)}_{A, \rm em} &=& Z_\pi \left[\frac{1 + 3 g^{(0) 2}_A}{2}
% \left(  \log \frac{\mu^2}{m_\pi^2} -1 \right) - g_A^{(0)2} \right]+ \hat{C}_{A}(\mu)^{(0),\,2} + \hat{C}_{\pi}(\mu)\right]
% \nonumber  \\
% \hat{C}_{A}(\mu) &=&
% 8 \pi^2 \left[ -  \frac{X_6}{2}  +
%  \frac{1}{g_A^{(0)}}  \left[ \tilde{X}_3 + \left( g_1 + g_2 + \frac{g_{11}}{2} \right) \right] \right] \  \  \
% \nonumber \\
%  \Delta^{(0)}_{V, \rm em} &=& \hat{C}_V =
% 8 \pi^2 \left[ -  \frac{X_6}{2}  +  2 \left( \tilde{X}_1 - \tilde{X}_2 \right) + g_9
%  \right]
%\end{eqnarray}
We provide in the Appendix
the explicit dependence of $\hat{C}_{V,A}$
on the LECs of $\mathcal{O}(e^2p)$.
Here we note that as written, $\hat{C}_{V,A}$ contain information about short-distance physics and in particular large logarithms connecting the weak scale to
the hadronic scale~\cite{Czarnecki:2004cw} and finite terms that have been calculated via dispersive
methods~\cite{Seng:2018yzq,Seng:2018qru,Czarnecki:2019mwq,Shiells:2020fqp}.

%We now turn to the NLO diagrams in the lower panel of  Fig.~\ref{Fig:DiagLO}. Diagrams $(2a)$ and $(2b)$ are reproduced by similar diagrams in $\slashpi$EFT, and do not contribute to the matching. Diagrams $(2c)$ to $(2f)$ vanish at $\mathcal O(\epsilon_\slashpi^0)$.  At $q=0$, all diagrams contributing to the vector current are cancelled by the WFR. We are left with a correction to $g_A$ that is dominated by diagrams $(2k)$ and $(2l)$.
%
%We now turn to the NLO diagrams in the lower panel of  Fig.~\ref{Fig:DiagLO}.
%
A similar analysis applies to the NLO amplitude, for which we report
a few representative diagrams in the lower panel of  Fig.~\ref{Fig:DiagLO}. % Diagrams $(2a)$ and $(2b)$ are reproduced by similar diagrams in $\slashpi$EFT, and do not contribute to the
% matching. Diagrams $(2c)$ to $(2f)$ vanish at $\mathcal O(\epsilon_\slashpi^0)$.
At $q=0$, all diagrams contributing to the vector operator are cancelled by the WFR, resulting in $\Delta^{(1)}_{V, \rm  em}=0$. We are left with a correction to $g_A$
\begin{eqnarray}
 \Delta^{(1)}_{A, \rm  em}
  = Z_\pi \,  4\pi m_\pi  \left[c_4 - c_3  + \frac{3}{8 m_N} + \frac{9}{16 m_N} g_A^{(0)2}\right]\,,
\end{eqnarray}
dominated by the LECs  $c_{3,4}$ from ${\cal L}_{\pi N}^{p^2}$
that contribute via topology (a2).
%are known LECs and [that is dominated by the $(a2)$ topology. \textcolor{red}{which part of the Eq.?}]

{\it Matching at $\mathcal O(\alpha  \epsilon_\slashpi)$ --- }
Through our final matching step, we identify additional isospin breaking terms to the LECs of the pion-less Lagrangian. Specifically, the pion loops with the vector current coupling to two pions (topology $(f1)$)
induce an isospin-breaking correction to the weak magnetism term.
In terms of the physical nucleon magnetic moments, $\mu_{n/p}$ (themselves
containing electromagnetic shifts), we find
%\begin{eqnarray}
%% \kappa_{\rm weak} = \kappa^{(0)}_p - \kappa^{(0)}_n  +  \frac{\alpha_{\rm em} Z_\pi}{2\pi} \frac{ g_A^2 m_N \pi}{ m_\pi}  \\
%% \kappa^{}_p - \kappa^{}_n = \kappa^{(0)}_p - \kappa^{(0)}_n  +  2 \frac{\alpha_{\rm em} Z_\pi}{2\pi} \frac{ g_A^2 m_N \pi}{ m_\pi}
%  \mu_{\rm weak} = \mu^{(0)}_p - \mu^{(0)}_n  +  \frac{\alpha_{\rm em} Z_\pi}{2\pi} \frac{ g_A^2 m_N \pi}{ m_\pi}  \\
% \mu^{}_p - \mu^{}_n = \mu^{(0)}_p - \mu^{(0)}_n  +  2 \frac{\alpha_{\rm em} Z_\pi}{2\pi} \frac{ g_A^2 m_N \pi}{ m_\pi}
%\end{eqnarray}
\begin{equation}\label{weak}
\mu_{\rm weak} - (\mu_p - \mu_n) =  -   \frac{\alpha Z_\pi}{2\pi} \frac{ g_A^2 m_N \pi}{ m_\pi}\,.
%\ = -3.3 \cdot 10^{-2},
\end{equation}
%corresponding to a $0.7\%$ correction.
which is not captured in experimental analyses thus far. Finally, the pion-$\gamma$ box  $(b1)$
%contributes to the tensor coupling to find
induces the tensor coupling
\begin{equation}\label{tensor}
 c_T =  \frac{\alpha}{2\pi} \frac{ g_A m_N \pi}{3 m_\pi}\,.
%= 0.011,
\end{equation}
%concluding our approach.
We discuss the numerical implications of these results  below.

\bigskip
{\it Connection to previous literature ---  } Recent approaches using current algebra and dispersion techniques \cite{Hayen:2020cxh, Gorchtein:2021fce} evaluated axial contributions as originating from vertex
corrections, in which the virtual photon is emitted and absorbed by the hadronic line,
and $\gamma W$ box, in which the virtual photon is exchanged between the hadronic and electron lines.
%and {\color{red} $\gamma W$ box} diagrams.
The latter was found to be largely consistent with the vector contribution using experimental data of the polarized Bjorken sum rule \cite{Hayen:2020cxh} and additional nucleon scattering data \cite{Gorchtein:2021fce}, as such including inelastic contributions without explicit calculation. The vertex corrections, on the other hand, have only been calculated in limiting scenarios. Following the notation of Ref. \cite{Hayen:2020cxh}, the \textit{a priori} non-zero contribution depends on a three-point function
\begin{align}
    \mathcal{D}_\gamma &= \int \frac{d^4k}{k^2}\int d^4y e^{i\bar{q}y}\int d^4x e^{ikx} \nonumber \\
    &\times \langle  p_f | T\left\{\partial_\mu J^\mu_W(y) J^\lambda_\gamma(x)J_\lambda^\gamma(0) \right\} | p_i \rangle\,,
    \label{eq:D_gamma}
\end{align}
where $\gamma (W)$ denotes electromagnetic (weak) currents, and $T\{\ldots\}$ the time-ordered product. In the chiral limit the divergence of the weak axial current vanishes ($\partial_\mu A^\mu \propto m_\pi \rightarrow 0$), while the vector current is conserved to higher order corrections in $\alpha$ and $m_d-m_u$.
%{\color{red} generally conserved up to higher-order corrections}.
Ref. \cite{Hayen:2020cxh} only considered the asymptotic and elastic contributions to Eq. (\ref{eq:D_gamma}), i.e. inserting a complete set of states in between every current and retaining only the nucleon. Assuming isospin symmetry then leads to a vanishing contribution
for the three-point function~\cite{Hayen:2020cxh}.
%recovering the result from Ref. \cite{Hayen:2020cxh}.
Recognizing diagrams $i1, j1, a2, \ldots$ in Fig. \ref{Fig:DiagLO} to correspond to an explicit treatment of these vertex corrections, the results presented here expand upon the simplified approach of Ref. \cite{Hayen:2020cxh} to find much larger than anticipated isospin-breaking corrections.

\bigskip
{\it Numerical impact ---}
We now estimate the numerical impact of the various corrections, starting with our main new finding, i.e.,  the  electromagnetic shift to $\lambda = g_A/g_V$. Including BSM contributions%new physics effects
, the relation between the experimentally extracted $\lambda$ and the (isosymmetric) QCD axial charge is given by~\cite{Bhattacharya:2011qm}
\begin{equation}
    \lambda = g_A^\mathrm{QCD} \Big(1 +\delta^{(\lambda)}_\mathrm{RC}  - 2\mathrm{Re}( \epsilon_R) \Big)~,
\end{equation}
where $\epsilon_R \sim (246\,\mathrm{GeV}/\Lambda_\mathrm{BSM})^2$ is a BSM right-handed current contribution appearing at an energy scale $\Lambda_\mathrm{BSM}$ \cite{Bhattacharya:2011qm,Alioli:2017ces}. To the order we are working the radiative correction is
\begin{equation}
\delta^{(\lambda)}_\mathrm{RC} =
\frac{\alpha}{2 \pi} \left( \Delta^{(0)}_{A, \rm em}  + \Delta^{(1)}_{A, \rm em}  - \Delta^{(0)}_{V \rm em}
\right)~.
%\lambda = g_A^{\rm QCD} \left(1 +  \Delta^{(0)}_{A, \rm em}  + \Delta^{(1)}_{A, \rm em}  - \Delta^{(0)}_{V \rm em} \right) ~.
\end{equation}
For the numerical evaluation of the loop contributions to  $\Delta^{(0),(1)}_{A, \rm em}$ we use $Z_\pi = 0.81$ (obtained from the physical pion mass difference and  $F_\pi = 92.4$~MeV) and  the average nucleon mass $m_N = 938.9$ MeV. In the loops we set $g_A^{(0)} = g_A \approx 1.27$~\cite{ParticleDataGroup:2020ssz}, as the difference formally contributes to higher chiral order. Existing lattice data indeed indicate that $g_A$ has a mild $m_\pi$ dependence~\cite{Chang:2018uxx,Gupta:2018qil}.
%%%%%% OLD  %%%%%%%%%%%
%Existing lattice data indicate that $g_A$ has a mild $m_\pi$ dependence \cite{Chang:2018uxx}, and we set $g_A^{(0)} = g_A = 1.27$.
%Using the physical masses of charged and neutral pions, the average nucleon mass $m_N = 938.9$ MeV, and
%$F_\pi = 92.4$ MeV, we obtain $Z_\pi = 0.81$.
%%%%%%%%%%%%%%%%%%%%%%
The NLO LECs $c_3$ and $c_4$ have been extracted from pion-nucleon scattering \cite{Hoferichter:2015hva,Siemens:2016jwj}.
They show a sizable dependence on the chiral order at which the fit to $\pi$-$N$ data is carried out,
with a big change between NLO and N$^2$LO, stabilizing between N$^2$LO and N$^3$LO.
 %%%%%%%%%%%%%%%%%%%%%%%%%%%%%%%%%%%
 %We use
%\begin{eqnarray}
%c_3 &=&-\{3.61(5),\,5.39(5),\,5.67(6)\}\, {\rm GeV}^{-1} \,,\nonumber \\
%c_4 &=&\phantom{-}\{2.17(3),\,3.62(3),\,4.35(4)\}\, {\rm GeV}^{-1}\,,
%\end{eqnarray}
%denoting NLO, N$^2$LO, and N$^3$LO extractions.
 %%%%%%%%%%%%%%%%%%%%%%%%%%%%%%%%%%%
%Using $c_{3,4}$ from Ref.~\cite{Siemens:2016jwj} we obtain
For the corrections we find
\begin{equation}\label{estimates}
\Delta^{(0)}_{A-V, \rm em} \in \{ 2.4, \, 5.7 \}\,, \
 \Delta^{(1)}_{A, \rm em} = \{ 10.0,    14.5,    15.9 \},
\end{equation}
where the range in $\Delta^{(0)}_{A-V, \rm em}$ is obtained by setting $\hat C_{A}(\mu) - \hat{C}_V=0$ and varying $\mu$ between $0.5$ and $1$ GeV,
while the three values of   $\Delta^{(1)}_{A, \rm em}$ are obtained by
using $c_{3,4}$ extracted to
NLO, N$^2$LO, and N$^3$LO~\cite{Siemens:2016jwj}.
While the NLO correction is somewhat larger than the LO one, we stress that
we do not know  the full LO correction because we have set
 the counter term contribution $\hat C_{A} - \hat{C}_V$  to zero. In addition, in an EFT without explicit $\Delta$ degrees of freedom, $c_3$ and $c_4$ are dominated by
 %virtual
 $\Delta$ contributions and thus anomalously large.
Combining the corrections, we estimate a correction to $\lambda$ at the percent level,
\begin{eqnarray}\label{gAshift}
%\delta g_A/g_A^{(0)}= \frac{\alpha}{2\pi} \Delta^{(0+1)}_{\rm em} &\in&  \{ 1.4, \,2.5 \} \cdot 10^{-2}\,.
\delta^{(\lambda)}_\mathrm{RC}
\in   \{ 1.4, \,2.6 \} \cdot 10^{-2}\,.
\end{eqnarray}
This shift has no impact on the current first-row CKM discrepancy because the most accurate determination of $\lambda$ is at present obtained from experiments, where these corrections are automatically included.
The correction does have a big impact when comparing with first-principles lattice QCD computations of neutron $\beta$ decay. Present lattice calculations of $g_A$ work in the isospin limit without QED, but Eq.~\eqref{gAshift} shows these results cannot be directly compared to the experimentally extracted value of $g_A$ without subtracting the newly identified isospin-breaking radiative corrections in this Letter.

%without a determination of the LECs entering $\hat C_{A} - \hat{C}_V$. \jdv{I was thinking to make this more direct and less technical by saying: ...cannot be directly compared to the experimentally extracted value of $g_A$ without subtracting the newly identified isospin-breaking corrections in this Letter.}
%Since the chiral expansion for $g_A$ in $\Delta$-less HB$\chi$PT is known to converge very slowly, with strong cancellations between different orders \cite{Bernard:2006te}, we cannot rule out that the same happens for electromagnetic corrections. Still, Eq. \eqref{estimates} indicates a potentially percent level correction,
%which needs to be investigated on the lattice.

\begin{figure}
    \centering
    \includegraphics[width=0.48\textwidth]{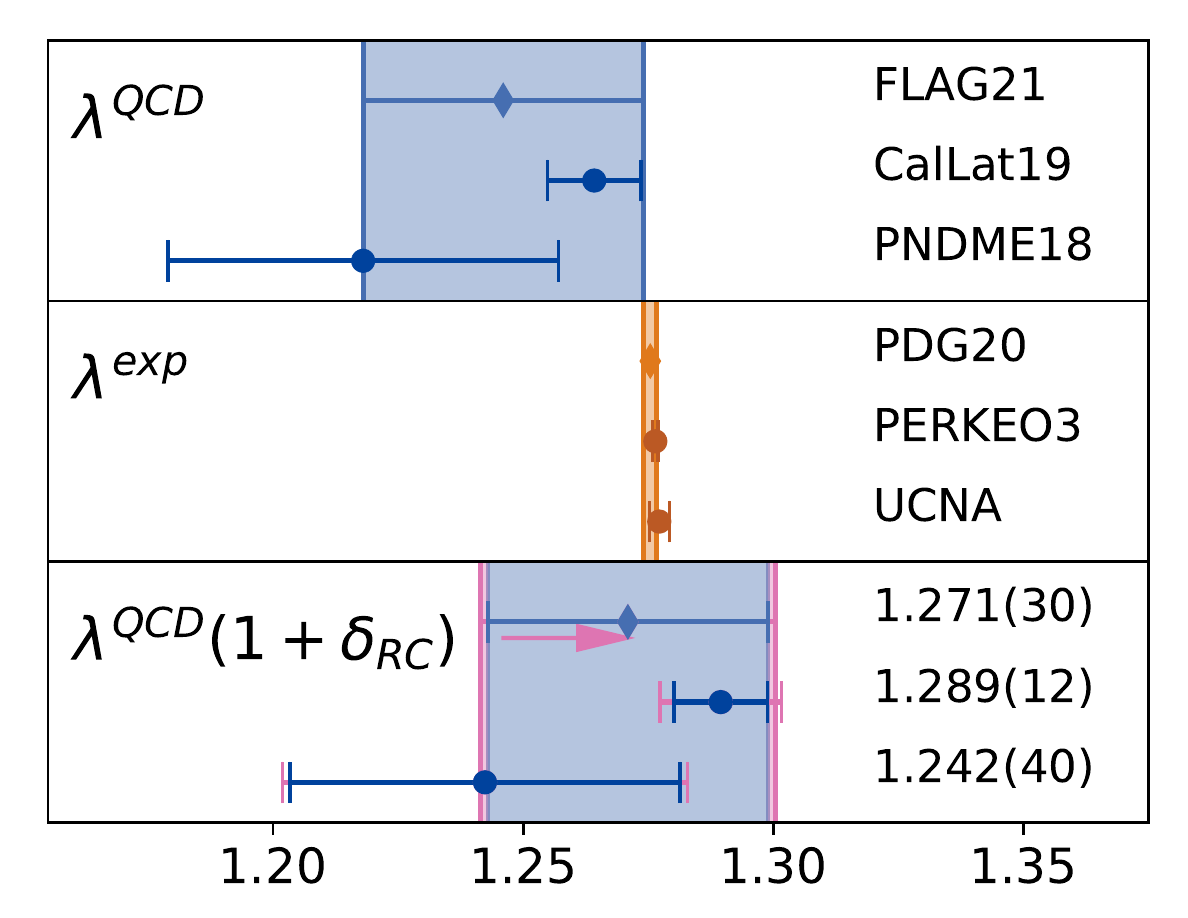}
    \caption{Overview of the required shift to lattice QCD determinations of $g_A$ and comparison with current experimental determination of $\lambda$. The bottom panel shows the shift and increased uncertainty in magenta with corrected values.  The keys in the figure are FLAG21~\cite{Aoki:2021kgd}, CalLat19~\cite{Walker-Loud:2019cif}, PNDME18~\cite{Gupta:2018qil}, PDG21~\cite{ParticleDataGroup:2020ssz}, PERKEO3~\cite{Markisch:2018ndu}, UCNA~\cite{UCNA:2017obv}.}
    \label{fig:gA_shift}
\end{figure}

%We show the significance of our main correction (Eq. (\ref{gAshift})) in Fig. \ref{fig:gA_shift} together with the state-of-the-art experimental determination of $\lambda$.

In Fig. \ref{fig:gA_shift}
we show the significance of the correction $\delta^{(\lambda)}_\mathrm{RC}$
in comparing lattice QCD calculations
with the state-of-the-art experimental determination of $\lambda$. %\jdv{Should we say somewhere that present lattice QCD computations do not include QED? [it is now said indirectly in the paragraph above - ``Present lattice calculations ... work in the isospin limit'']}
Compared to the most precise individual lattice calculation \cite{Walker-Loud:2019cif}, our radiative corrections corresponds to a 2.7$\sigma$ shift and a more modest $\sim 1 \sigma$ shift in the conservative FLAG'21 average \cite{Aoki:2021kgd}.
%%%%%%%%%%%%%%%%%%%%%%%%%%%%
%The comparison of the lattice and experimental determination of $\lambda$ is a promising channel for exotic right-handed interactions
%\begin{equation}
%    \lambda = g_A^\mathrm{QCD}\left\{1+\delta_\mathrm{RC}-2\mathrm{Re}( \epsilon_R)\right\}
%\end{equation}
%where $\epsilon_R \sim (80\,\mathrm{GeV}/\Lambda_\mathrm{BSM})^2$ is a BSM contribution appearing at an energy scale $\Lambda_\mathrm{BSM}$.
%%%%%%%%%%%%%%%%%%%%%%%%%%%%
%The radiative corrections
$\delta^{(\lambda)}_\mathrm{RC}$
generally improves the agreement between lattice QCD and experimental determination
of $\lambda$ and is essential if one wishes to obtain robust ranges (or constraints) on right-handed currents.
For example, assuming existing central values and an increased lattice-QCD precision,
the neglect of radiative corrections ($\delta^{(\lambda)}_\mathrm{RC}$) would  wrongfully point to BSM physics at
$\mathcal{O}(1\,\mathrm{TeV})$.

Isospin-breaking corrections to the weak magnetism do translate into explicit spectral changes (see the appendix for the full differential decay rate).
Relative corrections of $\mathcal{O}(10^{-4})$ occur in the SM predictions of both $a$, the $\beta$-$\nu$ angular correlation, and $A$, the $\beta$-asymmetry.
%A relative shift of \jdv{$2\cdot 10^{-4}$ occurs in the SM prediction of $a$,} the $\beta$-$\nu$ angular correlation, i.e.
These are comparable to anticipated experimental precision in the coming decade within the context of CKM unitarity tests~\cite{Cirgiliano:2019nyn}. Even larger relative changes ($\mathcal{O}(0.1\%)$) can occur due to cancellations in the leading-order SM prediction, such as in nuclear mirror systems used in complementary $|V_{ud}|$ determinations \cite{Naviliat-Cuncic:2008rcs}. An extension of this effort to nuclear systems is deemed crucial
and fits within rejuvenated superallowed efforts \cite{Gorchtein2018, Hardy:2020qwl}.
On the other hand, the induced tensor coupling $c_T$ produces a shift to the Fierz term and the
neutrino-asymmetry parameter $B$ at the level of $10^{-5}$,  negligible in light of the expected experimental accuracies.

%\bigskip
%\input{decayrate}

%\bigskip
%\input{lattice}

\bigskip
{\it Conclusions and outlook ---}
By using a systematic effective field theory approach we have identified and computed novel radiative corrections to neutron $\beta$-decay. %The corrections were argued to be small
%in previous literature based on erroneous power-counting arguments, but turn out to lead to sizable corrections.
The largest correction, at the percent level, can be understood as a QED correction to the nucleon axial charge. While this does not impact the extraction of ${\rm V}_{ud}$ from experiments,
%nor the associated tests of unitarity of the CKM quark mixing matrix,
it has important consequences for
the potential of $\beta$-decay experiments to constrain BSM right-handed currents when comparing the measured value of $\lambda=g_A/g_V$ to the first-principles calculation of the same quantity with lattice QCD.
%first-principle lattice-QCD computations of $\lambda=g_A/g_V$ and for the potential of $\beta$-decay experiments to constrain BSM right-handed currents. %~\cite{Bhattacharya:2011qm,Alioli:2017ces}.
In addition, we have identified changes in the neutron differential decay rate, in particular a shift in the $\beta$-$\nu$ angular correlation and the $\beta$-asymmetry, that are relevant for next-generation experiments.

%We have identified new, percent-level QED radiative  corrections to the neutron $\beta$-decay amplitude that have been previously missed in the literature.
%These corrections were determined with $SU(2)$ Heavy Baryon $\chi$PT that captures the nucleon structure corrections arising from virtual pion loops, and matched onto the lower-energy pion-less EFT.
%The large percent-level corrections can be understood as QED corrections to the nucleon axial charge, and therefore do not impact the extraction of ${\rm V}_{ud}$, but they do complicate the use of the experimental measurement and lattice QCD calculation of $\lambda=g_A/g_V$ to constrain BSM right-handed currents~\cite{Bhattacharya:2011qm,Alioli:2017ces}.

The new shift in the nucleon axial charge depends upon non-analytic contributions associated with pion loops as well as analytic short-distance corrections parameterized by LECs.
%the LECs of the theory that are needed to renormalize the amplitude.
The LECs that lead to the largest part of the correction ($c_3$ and $c_4$) are precisely extracted from pion-nucleon scattering data, but others are presently unknown leading to a sizable uncertainty in our results. %While the LECs contributing to NLO are well estimated in the literature ($c_3$ and $c_4$), those appearing at LO are not otherwise known or estimated, preventing us from making a full prediction for the radiative corrections.
Lattice QCD can compute the hadronic $n\rightarrow p$ amplitude in the presence of QED~\cite{Carrasco:2015xwa,Giusti:2017dwk}, which enables a determination of the unknown LECs.
There are subtleties that must be addressed related to gauge invariance and the non-factorizable contributions to the renormalization of the four-fermion operator~\cite{DiCarlo:2019knp}.
${\rm QED}_{\rm M}$~\cite{Endres:2015gda}, in which the photon is given a non-zero mass, may simplify the identification of the matrix element of interest by increasing the energy gap to the excited state contamination.

Looking beyond neutron decay, it is very possible that similar-sized corrections affect nuclear $\beta$-decay. The computations in this Letter provide the first step towards a full EFT treatment of radiative corrections to the multi-nucleon level. Given the interest in these low-energy precision tests of the Standard Model and the existing deviations from first-row CKM unitarity, it is imperative to accurately determine these radiative corrections in order to make full use of the anticipated precision of upcoming experiments.
%which can also be estimated through an EFT treatment.
 %Given the interest in these low-energy precision tests of the Standard Model, it is important that these radiative corrections be determined in order to make full use of the anticipated precision of upcoming experiments.
\\

\begin{acknowledgments}

\emph{Acknowledgements}.---We thank Misha Gorchtein and Martin Hoferichter for interesting conversations.
The work of AWL was supported in part by the U.S. Department of Energy, Office of Science, Office of Nuclear Physics under Awards No. DE-AC02-05CH11231.
EM is supported  by the US Department of Energy through
the Office of Nuclear Physics  and  the
LDRD program at Los Alamos National Laboratory. Los Alamos National Laboratory is operated by Triad National Security, LLC, for the National Nuclear Security Administration of U.S.\ Department of Energy (Contract No. 89233218CNA000001). JdV acknowledges support from the Dutch Research Council (NWO) in the form of a VIDI grant.
L.H. acknowledges support by the U.S. National Science Foundation (Grant No. PHY-1914133), U.S. Department of Energy (Grant No. DE-FG02-ER41042)

\end{acknowledgments}

\section{Appendix}
\label{sect:appendix}

{\it Effective Lagrangians and power counting --- }
We start from two-flavor QCD in presence of external sources
\bea
{\cal L} &=& {\cal L}_{\rm QCD} -  \ \bar{q}_R ({s + i p}) q_L  \ - \  \bar q_L ( {s -i p} ) q_R
\nonumber \\
 &+&  \bar q_L \gamma^\mu { l_\mu} q_L
\ + \  \bar q_R \gamma^\mu {r_\mu} q_R
\label{eq:SM1}
\eea
where $q^T = (u, d)$ and  $s(x), p(x), l_\mu(x), r_\mu(x)$
 can be written in terms of quark mass, Standard Model gauge fields,
and external classical fields $\bar{s}, \bar{p}, \bar{l}_\mu, \bar{r}_\mu$  as follows
\begin{subequations}
\bea
\chi & \equiv  & B_0 (s + i p) = B_0  (m_q + \bar{s} + i \bar{p})
\\
l_\mu &=& - e Q_L^{EM} A_\mu \, + \, Q_L^{W} J_\mu^{\rm lept} \,+ \, Q_L^{W \dagger } J_\mu^{\rm{lept}\dagger}
\, + \, \bar{l}_\mu
\qquad \ \
\\
r_\mu &=& - e Q_R^{EM} A_\mu  \, + \,  \bar{r}_\mu~.
\eea
\end{subequations}
Here  $B_0$ is a constant with dimension of mass,
 $m_q$ is the quark mass matrix,
$Q_L^{EM} = Q_R^{EM} = \mathrm{diag} (q_u, q_d)$  (with     $q_u=2/3, q_d= -1/3$),
$Q_L^{W} =  - 2 \sqrt{2} G_F  V_{ud} \, \tau^+$, and  $ J_\mu^{\rm lept}  = \bar e_L \gamma_\mu \nu_{eL} $.
The Lagrangian in \eqref{eq:SM1} is invariant under local $G=SU(2)_L \times SU(2)_R \times U(1)_V$ transformations
\be
q_L \to L (x)e^{\alpha_V(x)}  q_L~, \quad
q_R \to R(x) e^{\alpha_V(x)}  q_R~,
\ee
with $L,R \in SU(2)_{L,R}$, provided
 $Q_{L,R}^{EM}$ and  $Q_L^W$ transform  as  ``spurions"  under the chiral group
$Q_L^{EM, W} \to L Q_L^{EM,W} L^\dagger$ and $Q_R^{EM} \to R Q_R^{EM} R^\dagger$,
and that $\bar{l}_\mu$ and $\bar{r}_\mu$ transform as gauge fields under  $G$.
This implies
%%%%%%%%%%%%%%%%%%%%%%
%In order to implement the effect of virtual photons in the low-energy chiral Lagrangian,
%it is convenient to treat $Q_{L,R}^{EM}$ and  $Q_L^W$ as  ``spurions", transforming under the chiral group
%($q_L \to L q_L$,
%$q_R \to R q_R$ with $L,R \in SU(2)_{L,R}$)  as
% $Q_L^{EM, W} \to L Q_L^{EM,W} L^\dagger$ and $Q_R^{EM} \to R Q_R^{EM} R^\dagger$.
%%%%%%%%%%%%%%%%%%%%%%
\begin{subequations}
\bea
\chi &\to & R \chi L^\dagger
\\
l_\mu &\to &  L  l_\mu L^\dagger + i L \partial_\mu L^\dagger + \partial_\mu \alpha_V
\\
r_\mu &\to&  R r_\mu R^\dagger + i R \partial_\mu R^\dagger + \partial_\mu \alpha_V~.
\eea
\end{subequations}
Note that the external sources can be decomposed in $SU(2)$ singlet and non-singlet components
as follows: $l_\mu = l_\mu^{ns} + l_\mu^s$, $r_\mu =  r_\mu^{ns} + r_\mu^s$.

To construct the effective chiral Lagrangians, one introduces the nucleon and pion fields as follows~\cite{Coleman:1969sm,Callan:1969sn},
\be
N = \left(
\begin{array}{c}
p
\\
n
\end{array}
\right),
\
%\qquad
%u = e^{i \Pi/(2 F)}
U = u^2  = e^{i \Pi/(F)},
\
%\qquad
\Pi = \left(
\begin{array}{cc}
\pi^0 & \sqrt{2} \pi^+
\\
\sqrt{2} \pi^- & - \pi^0
\end{array}
\right)
% \qquad U = u^2~.
\ee
 and $F \sim F_\pi = 92.4$~MeV.
 These fields transform under the chiral group as follows
 \begin{subequations}
 \bea
 u  & \to  & L u K^\dagger (u) = K (u) u R^\dagger
 \\
 U & \to & L U R^\dagger
 \\
 N & \to & e^{3 i \alpha_V}  K(u) N
 \eea
 \end{subequations}
where   $K(u)$ is a pion-dependent $SU(2)_V$ transformation.

 To construct chiral invariant Lagrangians, it is very useful to use chiral-covariant derivatives
 \begin{subequations}
 \bea
 D_\mu U  & \equiv  & \partial _\mu U - i l_\mu U + i U r_\mu \to L (D_\mu U) R^\dagger
 \\
 \nabla_\mu N &\equiv & \left(\partial_\mu + \Gamma_\mu - i \frac{3(l_\mu^s + r_\mu^s)}{2}  \right) N \to K (\nabla_\mu N)
\quad \quad \
 \\
\Gamma_\mu &=& \frac{1}{2} \left[ u (\partial_\mu - i r^{ns}_\mu) u^\dagger + u^\dagger (\partial_\mu - i l^{ns}_\mu) u \right]
\nonumber \\
& \to &  K(u) \Gamma_\mu K(u)^\dagger + K (u) \partial_\mu K(u)^\dagger
 ~.
 \eea
 \end{subequations}

 It is also very useful to use combinations of fields that transform homogeneously with $K(u)$:
 \begin{subequations}
 \bea
u_\mu &=&  i \ \left[ u (\partial_\mu - i r_\mu) u^\dagger - u^\dagger (\partial_\mu - i l_\mu) u \right]
\nonumber \\
& \to &  K(u) u_\mu K(u)^\dagger
\\
\chi_\pm &=& u^\dagger \chi u^\dagger \pm u \chi^\dagger u \to  K(u)  \chi_\pm  K(u)^\dagger
\\
{\cal Q}_L^{EM,W} &=& u^\dagger Q_L^{EM,W} u \to  K(u) {\cal Q}_L^{EM,W} K(u)^\dagger
\\
{\cal Q}_R^{EM} &=& u Q_R^{EM} u^\dagger  \to  K(u) {\cal Q}_R^{EM} K(u)^\dagger
 \eea
 \end{subequations}
Finally,  in the literature one often finds the combinations of charge building blocks with definite parity
\be
{\cal Q}_\pm \equiv \frac{1}{2} \left( {\cal Q}_L \pm {\cal Q}_R \right)~.
\ee

%\subsection{Power counting}

The standard $\chi$PT power counting assumes that  external momenta and meson masses are comparable ($ q_{\rm ext} \sim m_\pi$).
Including charged lepton masses one assumes
%\be
$ p \sim q_{\rm ext}  \sim m_{\mu} \sim  m_\pi \ll \Lambda_\chi \sim 4 \pi F_\pi \sim m_N$.
%\ee
Given this, one  makes the following assignments:
\be
\partial \sim p~,
\quad
\chi_\pm \sim B_0 m_q \sim m_\pi^2 \sim p^2~.
%\qquad
%m_\ell \sim p~,
\quad
l_\mu, r_\mu \sim p~,
\ee
with the latter identification implying $e \sim p$ and $G_F \sim p$ (though we will never go beyond one insertion of $G_F$ and two insertions of the
electromagnetic coupling $e$). The above scalings allow us to assign chiral dimension to each lagrangian vertex in a straightforward way.

The pion Lagrangian has the usual expansion in even chiral powers:
\begin{subequations}
\bea
{\cal L}_\pi &=& {\cal L}_\pi^{(2)} +  {\cal L}_\pi^{(4)} + ...
\\
{\cal L}_\pi^{(2)} &=&
{\cal L}_\pi^{p^2}  +
{\cal L}_\pi^{e^2 p^0}
\nonumber \\
&=&
\frac{F^2}{4} \langle u_\mu u^\mu + \chi_+  \rangle  + e^2 Z_\pi F^4 \langle {\cal Q}_L^{EM} {\cal Q}_R^{EM}  \rangle,  \qquad
%\\
%{\cal L}_\pi^{(4)} &=& ...
\eea
\end{subequations}
which leads to the identification
\be
m_{\pi^\pm}^2 - m_{\pi^0}^2 = 2 e^2  F_\pi^2 Z_\pi ~.
\ee

The gauge-kinetic leptonic Lagrangian has chiral dimension $n=1$:
\be
{\cal L}_{\rm lept} = \bar e \left( i  \slashed{\partial} + e \slashed{A} - m_e \right) e  + \bar \nu  i \slashed{\partial} \nu~.
\ee
To the order we work, we need to include the following purely leptonic counter-term~\cite{Knecht:1999ag}
\be
{\cal L}_{\rm lept}^{CT} = e^2 X_6 \ \bar e \left( i  \slashed{\partial} + e \slashed{A}  \right) e  ~.
\ee

The pion-nucleon Lagrangian has both odd and even chiral powers, starting at $n=1$:
\begin{subequations}
\bea
{\cal L}_{\pi N} &=& {\cal L}_{\pi N}^{(1)} +  {\cal L}_{\pi N}^{(2)} +   {\cal L}_{\pi N}^{(3)} + ...
\\
{\cal L}_{\pi N}^{(1)} &=&
{\cal L}_{\pi N}^{p} =
\bar N_v i v \cdot \nabla N_v + g_A \bar N_v S \cdot u N_v
\\
{\cal L}_{\pi N}^{(2)} &=&   {\cal L}_{\pi N}^{p^2} + {\cal L}_{\pi N}^{e^2 p^0}
\\
{\cal L}_{\pi N}^{(3)} &=&   {\cal L}_{\pi N}^{p^3} + {\cal L}_{\pi N}^{e^2 p} + {\cal L}_{\pi N \ell}^{e^2 p}
\eea
\end{subequations}
where in the nucleon rest-frame
$v^\mu = (1, \bf 0)$ and $S^\mu = (0, \boldsymbol{\sigma}/2)$.
We have displayed  explicitly here only the leading order Lagrangians and we will report below  the appropriate higher order terms
as needed.
All these effective Lagrangian are know in the literature, see for example Ref.~\cite{Gasser:2002am},
except for ${\cal L}_{\pi N \ell}^{e^2 p}$, which is needed to reabsorb divergences from loops that involve virtual baryons, pions, leptons, and photons.
We report here only the terms that play a significant role in our analysis.

The one-loop diagrams with virtual nucleons, pions, and photons generate divergences which are absorbed by counterterms in the ${\cal L}_{\pi N}^{e^2 p}$ Lagrangian.
When constructing the baryon electromagnetic Lagrangian,
it has been  common practice in the literature~\cite{Meissner:1997ii,Muller:1999ww,Gasser:2002am}  to  use
charge spurions corresponding to the nucleon charge matrix $\bar Q = diag (1,0)$.
Now $\bar Q$   differs from the quark charge matrix only in its $SU(2)$ singlet component:
 therefore the two objects  have the same transformation properties under the chiral group and this procedure
 is justified. In what follows we indicate all the chiral building blocks built from the nucleon charge matrix with a bar.
A minimal version of ${\cal L}_{\pi N}^{e^2 p}$  was constructed in Ref.  \cite{Gasser:2002am}
\begin{equation}
{\cal L}_{\pi N}^{e^2 p}  =  e^2    \sum_{i=1,12} \  g_i \ \bar N_v \, O_i^{e^2 p} \, N_v ~,
\end{equation}
Only four operators contribute to neutron decay at tree level,
\begin{subequations}
\begin{eqnarray}
%O_1^{e^2 p} &=&  \langle {\cal Q}_+^2 - {\cal Q}_-^2 \rangle \, S \cdot u
%\\
%O_2^{e^2 p} &=&  \langle {\cal Q}_+ \rangle^2  \, S \cdot u
%\\
%O_9^{e^2 p} &=&  \frac{i}{2}  [ {\cal Q}_+,  v \cdot c^+ ] + {\rm h.c.}
%\\
%O_{11}^{e^2 p} &=&   \frac{i}{2}  [ {\cal Q}_+,  S \cdot c^- ]  ~,
%
O_1^{e^2 p} &=&  \langle {\cal \bar Q}_+^2 - {\cal \bar Q}_-^2 \rangle \, S \cdot u
\\
O_2^{e^2 p} &=&  \langle {\cal \bar Q}_+ \rangle^2  \, S \cdot u
\\
O_9^{e^2 p} &=&  \frac{i}{2}  [ {\cal \bar Q}_+,  v \cdot c^+ ] + {\rm h.c.}
\\
O_{11}^{e^2 p} &=&   \frac{i}{2}  [ {\cal \bar Q}_+,  S \cdot c^- ]  ~,
\end{eqnarray}
\end{subequations}
with
\be
c_\mu^\pm = - \frac{i}{2}  \left(  u [l_\mu, \bar Q] u^\dagger \pm u^\dagger [r_\mu,\bar Q] u \right)  ~.
%\qquad  r_\mu =0~, \quad l_\mu = - 2 \sqrt{2} G_F V_{ud} \bar{e}_L \gamma_\mu \nu_L \ \tau^+ ~.
\ee

As standard practice in $\chi$PT, the divergences are subtracted as follows~\cite{Gasser:1983yg}:
\bea
 g_i &=& \eta_i  \ \lambda (\mu) +  g_i^r(\mu)~,
 \nonumber \\ \label{eq:msbar_gl}
 \lambda (\mu) &=& \frac{\mu^{d-4}}{(4 \pi)^2} \left( \frac{1}{d-4} - \frac{1}{2} \left( -\gamma + \log 4 \pi  + 1 \right) \right).
\eea
We use the same subtraction scheme for all LECs.
The coefficients  $\eta_i$ can be found in Table 5 of Ref.~\cite{Gasser:2002am}.
We checked that the $g_i$ couplings  absorb correctly the divergences of diagrams without virtual leptons,
thus providing a consistency check on our calculation.

The one-loop diagrams with virtual nucleons, pions,  photons, and charged leptons generate divergences which are absorbed by counterterms in the
new  ${\cal L}_{\pi N \ell }^{e^2 p}$ Lagrangian.
These are the analogue of the operators introduced in the meson sector  in Ref.~\cite{Knecht:1999ag}, that contribute to (semi)leptonic meson decays to $O(e^2 p^2)$.
We find five structures, of which only the first three contribute to neutron decay at tree level
\be
%{\cal L}_{\pi N \ell}^{e^2 G_F} =
{\cal L}_{\pi N \ell}^{e^2 p} =
e^2 \sum_{i=1,5}  \ \tilde X_i \, \tilde{O}_i ~,
\ee
where
\begin{subequations}
\label{eq:basis}
\bea
\tilde  O_1 &=&   \bar e \gamma_\alpha \nu_L  \ \bar N_v v^\alpha {\cal Q}_L^W N_v
\\
\tilde  O_2 &=& \bar e \gamma_\alpha \nu_L  \ \bar N_v v^\alpha  [{\cal Q}_L^W,  {\cal \bar Q}_R^{EM} ]  N_v
\\
\tilde  O_3 &=& \bar e \gamma_\alpha \nu_L  \ \bar N_v S^\alpha  [{\cal Q}_L^W,  {\cal \bar Q}_R^{EM} ]  N_v
\\
\tilde  O_4 &=&  \bar e \gamma_\alpha \nu_L  \ \bar N_v v^\alpha  \langle {\cal Q}_L^W  {\cal \bar  Q}_R^{EM} \rangle  N_v
\\
\tilde  O_5 &=& \bar e \gamma_\alpha \nu_L  \ \bar N_v S^\alpha  \langle {\cal Q}_L^W  {\cal  \bar Q}_R^{EM} \rangle  N_v~.
\eea
\end{subequations}
%Here $\langle ... \rangle$ denotes the trace in flavor space and the ${\cal Q}_{L,R}^{EM, W}$   are given by
%\bea
%{\cal Q}_L^{W,EM} &=& u^\dagger  Q_L^{W,EM} u
%\\
%{\cal Q}_R^{EM} &=& u  Q_R^{EM} u^\dagger~
%\eea
%and  $Q_{L,R}^{EM}$, $Q_L^W$ are the electromagnetic and weak spurions that eventually take the form:
%\be
%Q_{L,R}^{EM} =
%\left(
%\begin{array}{cc}
%1 & 0
%\\
%0 & 0
%\end{array}
%\right) \qquad  \qquad
%Q_L^W = - 2 \sqrt{2} G_F V_{ud}  \ \tau^+~.
%\ee
The couplings $\tilde X_i$ are dimensionless (note that ${\cal Q}_W$ carries dimension via the $G_F$ factor).

To compute the neutron decay amplitude to $O(G_F \alpha \epsilon_\chi)$ we must consider one-loop diagrams
with insertions of ${\cal L}_{\pi N }^{p^2}$, for which  (in the notation of  Ref. \cite{Bernard:1995dp})  we use
\begin{widetext}
\begin{eqnarray}
    \mathcal L^{p^2}_{\pi N} &=& \bar N \Bigg[ \frac{1}{2 m_N} \left( (v \cdot \mathcal D)^2 - \mathcal D^2 \right) - i \frac{g_A}{2 m_N} \left\{ S \cdot \mathcal D, v \cdot u\right\} + c_1 \textrm{Tr}(\chi_+)
    +  \left(c_2 - \frac{g^2_A}{8 m_N}\right) (v\cdot u)^2 \nonumber \\
    & & + c_3 u \cdot u + \left(c_4 + \frac{1}{4 m_N} \right) \left[S^\mu, S^\nu\right] u_\mu u_\nu  + c_5 \tilde\chi_+
%    \nonumber \\ & &
    - \frac{i}{4 m_N} \left[S^\mu,S^\nu\right]
    \left( (1+\kappa_1) f^+_{\mu\nu} + \frac{1}{2} (\kappa_0 - \kappa_1) \textrm{Tr}\left(f^+_{\mu\nu}\right)\right)
    \Bigg] N .  \qquad
     \label{Eq:LagPiN_nlo}
\end{eqnarray}
\end{widetext}

Given these Lagrangians, Weinberg's power counting argument~\cite{Weinberg:1978kz,Weinberg:1990rz,Weinberg:1991um} implies that
connected diagrams scale as ${\cal A} \sim p^\nu$
with
\be
\nu = 2 L + 1 + \sum_{n=2, 4,...} (n-2) N_n^M + \sum_{m=1,2,...} (m-1) N_m^F
\label{eq:pc1}
\ee
where $L$ is the number of loops
%$n$  ($m$) is the chiral dimension of vertices involving only mesons (mesons and baryons/leptons)
and $N_n^M$ ($N_m^F$) is the number of  mesonic (fermionic) vertices with chiral dimension $n$ ($m$).
In deriving this formula, pion propagators are counted as $p^{-2}$ and baryon / lepton propagators are counted as $p^{-1}$.

Using this power counting one sees that the amplitude for neutron decay can be organized as follows
\begin{subequations}
\bea
{\cal A} &=& {\cal A}^{(1)} +  {\cal A}^{(2)} +
{\cal A}^{(3)} +    {\cal A}^{(4)} +  ...
\\
{\cal A}^{(1)} &=&  {\cal A}^{G_Fp^0}
\\
{\cal A}^{(2)} &=&  {\cal A}^{G_Fp}
\\
{\cal A}^{(3)} &=&  {\cal A}^{G_Fp^2}  + {\cal A}^{e^2 G_Fp^0}
\\
{\cal A}^{(4)} &=&  {\cal A}^{G_Fp^3}  + {\cal A}^{e^2 G_Fp}
\\
&...&
\nonumber
\eea
\end{subequations}
We are interested in computing the leading  and next-to-leading  electromagnatic  corrections to the neutron decay, which appear at chiral order $n=3$ and $n=4$, respectively.
Using Eq.~(\ref{eq:pc1}) one can easily identify the tree-level and one-loop diagrams that contribute to a given order, up to
${\cal A}^{(4)}$:
\begin{itemize}
\item The amplitude ${\cal A}^{(1)}$
 is given by a tree-level diagram with insertion of the weak vertices from ${\cal L}_{\pi N}^{(1)}$.

\item  The amplitude ${\cal A}^{(2)}$ is  obtained by tree-level graphs with one insertion of ${\cal L}_{\pi N }^{(2)}$
and any number of insertions   from  ${\cal L}_{\pi N}^{(1)}$  and ${\cal L}_{\pi}^{(2)}$.
It contributes to neutron decay at order $G_F \epsilon_{\rm recoil}$.

\item  The amplitude ${\cal A}^{(3)}$ is given  by tree-level graphs with one insertion of ${\cal L}_{\pi N}^{(3)}$
and any number of insertions   from  ${\cal L}_{\pi N}^{(1)}$  and ${\cal L}_{\pi }^{(2)}$;
and by  one-loop diagrams with vertices from ${\cal L}_{\pi N}^{(1)}$  and ${\cal L}_{\pi }^{(2)}$.
In Fig. ~\ref{Fig:DiagLO}  (upper panel)  we show all
one-loop topologies contributing up to $O(e^2 G_Fp^0)$,
These involve virtual pions, virtual photons, and virtual charged leptons.

\item  The amplitude ${\cal A}^{(4)}$ is given  by tree-level graphs with one insertion of ${\cal L}_{\pi N}^{(4)}$
and any number of insertions   from  ${\cal L}_{\pi N}^{(1)}$  and ${\cal L}_{\pi }^{(2)}$;
and by one-loop diagrams with
one vertex from ${\cal L}_{\pi N}^{(2)}$   and any number of  vertices from ${\cal L}_{\pi N}^{(1)}$   and ${\cal L}_{\pi }^{(2)}$.
Note that  tree level graphs with insertion of ${\cal L}_{\pi N}^{(e^2 p^2)}$ do not contribute.
In Fig. ~\ref{Fig:DiagLO}  (lower panel)  we show representative
one-loop diagrams contributing up to $O(e^2 G_F p)$,
These involve virtual pions, virtual photons, and virtual charged leptons.

\end{itemize}

The counterterm contributions to the amplitude at $O(e^2 G_F p^0)$ are captured by the combinations
$\hat{C}_{V/A}$ (see Eq.~(\ref{eq:dAem0})) as follows:
\begin{eqnarray}
 \hat{C}_{A} &=&
 8 \pi^2 \left[ -  \frac{X_6}{2}  +
  \frac{1}{g_A^{(0)}}  \left[ \tilde{X}_3 + \left( g_1 + g_2 + \frac{g_{11}}{2} \right) \right] \right] \  \  \
   \nonumber \\
 \hat{C}_V &=&
 8 \pi^2 \left[ -  \frac{X_6}{2}  +  2 \left( \tilde{X}_1 - \tilde{X}_2 \right) + g_9
  \right]~.
\label{eq:CVALECs}
\end{eqnarray}

{\it Neutron decay rate --- } We now present the neutron differential decay rate up-to-and-including $\mathcal O(G_F \epsilon_{\mathrm{recoil}}$, $\mathcal O(G_F \alpha)$, $\mathcal O(G_F \alpha \epsilon_\chi)$, and $\mathcal O(G_F \alpha\epsilon_{\slashpi})$ corrections. We follow Refs.~\cite{Ando:2004rk, Gudkov:2005bu, Bhattacharya:2011qm} and write
\begin{equation}
\frac {d\Gamma}{d E_e d\Omega_e d \Omega_\nu} = \frac{ (G_F V_{ud})^2}{(2\pi)^5}(1 + 3\lambda^2) w(E_e) D(E_e, \vec p_e, \vec p_{\nu}, \vec \sigma_n)\,,
\end{equation}
where $\vec \sigma_n$ denotes the neutron polarization and $\lambda = g_A/g_V$. The spectrum of the electron is described by
\begin{equation}
w =  |\vec p_e| E_e (E_0-E_e)^2 F(E_e) \left(1+ \frac{\alpha}{2\pi}\delta^{(1)}_{\alpha}(E_e)\right)\,,
%\nonumber
%\left(1+\frac{\alpha}{2\pi} e_V^R + \frac{\alpha}{2\pi}\delta^{(1)}_{\alpha}(E_e)\right)\,,\nonumber
\end{equation}
where $E_0 = (m_n^2-m_p^2+m_e^2)/(2m_n)$ is the maximal electron energy, and $F(E_e)$ is the Fermi function for an electron in the field of the final-state proton. The radiative correction $\delta^{(1)}_{\alpha}$ is discussed below. The function $D$ can be parametrized\footnote{A possible correction to the time-reversal-odd $D$ coefficient only enters at $\mathcal O(G_F \alpha\epsilon_{\mathrm{recoil}})$ \cite{Ando:2009jk}.} as
\begin{eqnarray}
D &=& 1+ c_0 + c_1\frac{E_e}{m_N} + \frac{m_e}{E_e}\bar b + \bar a \frac{\vec p_e \cdot \vec p_\nu}{E_e E_\nu} + \bar A \frac{\vec\sigma \cdot \vec p_e}{E_e} \nonumber\\ &&+ \bar B \frac{\vec \sigma \cdot \vec p_\nu}{E_\nu}
+ \bar C_{aa} \left(\frac{\vec p_e \cdot \vec p_\nu}{E_e E_\nu} \right)^2\nonumber\\&&+ \bar C_{aA} \frac{\vec p_e \cdot \vec p_\nu}{E_e E_\nu} \frac{\vec\sigma \cdot \vec p_e}{E_e} + \bar C_{aB} \frac{\vec p_e \cdot \vec p_\nu}{E_e E_\nu} \frac{\vec \sigma \cdot \vec p_\nu}{E_\nu}\,.
\end{eqnarray}
The various coefficients can be further decomposed through \cite{Gudkov:2005bu,Bhattacharya:2011qm}
\begin{eqnarray}
\bar a &=& \left(a_{\mathrm{LO}}+c_0^{(a)}+c_1^{(a)}\frac{E_e}{m_N}\right)\left(1+\frac{\alpha}{2\pi}\delta_{\alpha}^{(2)}(E_e)\right)\,,\nonumber\\
\bar A &=& \left(A_{\mathrm{LO}}+c_0^{(A)}+c_1^{(A)}\frac{E_e}{m_N}\right)\left(1+\frac{\alpha}{2\pi}\delta_{\alpha}^{(2)}(E_e)\right),\nonumber\\
\bar B &=& B_{\mathrm{LO}}+c_0^{(B)}+c_1^{(B)}\frac{E_e}{m_N}+\frac{m_e}{E_e} b_\nu\,,\nonumber\\
\bar C_{aa} &=& c_1^{(aa)}\frac{E_e}{m_N}\,,\nonumber\\
\bar C_{aA} &=& c_1^{(aA)}\frac{E_e}{m_N}\,,\nonumber\\
\bar C_{aB} &=& c_0^{(aB)}+c_1^{(aB)}\frac{E_e}{m_N}\,.
\end{eqnarray}
The LO coefficients are well known and given by
\begin{eqnarray}
a_{\mathrm{LO}} &=& \frac{1-\lambda^2}{1+3 \lambda^2}\,,\nonumber \\
A_{\mathrm{LO}} &=& \frac{2 \lambda-2 \lambda^2}{1+3 \lambda^2}\,,\nonumber \\
B_{\mathrm{LO}} &=& \frac{2 \lambda+ 2 \lambda^2}{1+3 \lambda^2}\,.
\end{eqnarray}
We write the remaining coefficients in terms of $\bar \mu_V = \mu_p - \mu_n - \frac{\alpha Z_\pi}{2\pi}\frac{g_A^2 m_N \pi}{m_\pi}$ and $c_T = \frac{\alpha}{2\pi}\frac{g_A m_N \pi}{3m_\pi}$
\begin{eqnarray}
\bar b &=& -\frac{m_e}{m_N}\frac{1 + 2 \lambda( \bar \mu_V+ 3 c_T) + \lambda^2}{1+3 \lambda^2}\,,\nonumber\\
c_0 &=& -\frac{E_0}{m_N}\frac{ 2 \lambda( \lambda+\bar \mu_V) }{1+3 \lambda^2}\,,\nonumber\\
c_1 &=& \frac{ 3+4 \lambda \bar \mu_V+ 9 \lambda^2 }{1+3 \lambda^2}\,,\nonumber\\
c^{(a)}_0 &=& \frac{E_0}{m_N}\frac{ 2 \lambda( \lambda+\bar \mu_V) }{1+3 \lambda^2}\,,\nonumber\\
c^{(a)}_1 &=&- \frac{ 4 \lambda(3 \lambda+ \bar \mu_V) }{1+3 \lambda^2}\,,\nonumber\\
c^{(A)}_0 &=& \frac{E_0}{m_N}\frac{( \lambda -1)( \lambda+\bar \mu_V) }{1+3 \lambda^2}\,,\nonumber\\
c^{(A)}_1 &=& \frac{  \lambda(7 - 5 \lambda) + \bar \mu_V(1-3 \lambda) }{1+3 \lambda^2}\,,\nonumber\\
c^{(B)}_0 &=&- \frac{E_0}{m_N}\frac{ 2 \lambda( \lambda+\bar \mu_V) }{1+3 \lambda^2}\,,\nonumber\\
c^{(B)}_1 &=& \frac{ \lambda(7 + 5 \lambda) + \bar \mu_V(1 + 3 \lambda)}{1+3 \lambda^2}\,,\nonumber\\
b_\nu &=& -\frac{m_e}{m_N}\frac{(1 +  \lambda)(\lambda +  \bar \mu_V) +  2 c_T(1+2 \lambda)}{1+3 \lambda^2}\,,\nonumber\\
c^{(aa)}_1 &=&- \frac{ 3 (1- \lambda^2) }{1+3 \lambda^2}\,,\nonumber\\
c^{(aA)}_1 &=&\frac{ (\lambda-1)(5 \lambda + \bar \mu_V) }{1+3 \lambda^2}\,,\nonumber\\
c^{(aB)}_0 &=& \frac{E_0}{m_N}\frac{( 1+ \lambda )( \lambda+\bar \mu_V) }{1+3 \lambda^2}\,,\nonumber\\
c^{(aB)}_1 &=&- \frac{ (1+\lambda)(7 \lambda + \bar \mu_V)}{1+3 \lambda^2}\,.
\end{eqnarray}

The explicit expressions for the radiative corrections $\delta^{(1)}_{\alpha}$ and $\delta^{(2)}_{\alpha}$ are given by
%(we work in dimensional regularization with $d=4 -2 \epsilon$ and $\mu_{\overline{\rm MS}} =\mu \sqrt{4 \pi} e^{- \gamma_E}$)
\bea
\delta_\alpha^{(1)}
&=& 2\hat C_V +  %\left(
    %\frac{3}{2 \epsilon}
    %+ 3 \log \frac{\mu_{\overline{\rm MS}}}{m_e}  + 2 \right)
    3 \log \frac{\mu}{m_e}  + \frac{1}{2}
    %\right)
\nonumber \\
&&+ \frac{1 + \beta^2}{\beta} \, \log \frac{1+\beta}{1-\beta}
+ \frac{1}{12 \beta} \left( \frac{\bar E}{E_e}\right)^2  \log \frac{1+\beta}{1-\beta}
\nonumber \\
&&+4 \left[ \frac{1}{2 \beta} \log \frac{1+\beta}{1-\beta} - 1 \right] \left[ \log \frac{2 \bar E}{m_e}  - \frac{3}{2} + \frac{\bar E}{3 E_e} \right]
\nonumber \\
&&+
\frac{1}{\beta} \left[ - 4 \,  {\rm Li}_2 \left( \frac{2 \beta}{1 + \beta} \right) - \log^2 \left( \frac{1+\beta}{1-\beta} \right)   \right]\,,\\
\delta_\alpha^{(2)}
&=&  \frac{1- \beta^2}{ \beta}  \, \log \frac{1+\beta}{1-\beta}\nonumber \\
&&+ \frac{4 (1 - \beta^2)}{3 \beta^2} \frac{\bar E}{E_e}    \left[ \frac{1}{2 \beta} \log \frac{1+\beta}{1-\beta} -1 \right]
\nonumber \\
&+& \frac{1}{6 \beta^2}  \frac{\bar E^2}{E_e^2}   \left[ \frac{1 - \beta^2}{2 \beta} \log \frac{1+\beta}{1-\beta} -1 \right]\,,
\eea
where $\beta = |\vec p_e|/E_e$ and $\bar E = E_0 - E_e$.
Our expression for $\delta_\alpha^{(1)}$
coincides with the combination of $\delta_\alpha^{(1)} + e_V^R (\mu)$ in Ref.~\cite{Ando:2004rk}
upon identifying $2 \hat{C}_V$ with the combination  of counterterms $e_V - (e_1 + e_2)/2$ ] in Ref.~\cite{Ando:2004rk}.
Finally, expressing $\delta_{\alpha}^{(1)}$ in terms of the Sirlin function $g(E_e, E_0)$~\cite{Sirlin:1967zza}, we find
\begin{equation}
\delta_{\alpha}^{(1)} = %\frac{3}{2\varepsilon}
    2 \hat C_V + \frac{5}{4} + 3 \log \frac{\mu}{m_p} + g(E_e, E_0)\,.
\end{equation}

\bibliography{bibliography}

\end{document}